\documentclass[12pt,graphicx,subfigure,axodraw]{article}
\usepackage{amsfonts}
\usepackage{amssymb}
\usepackage{epsfig}
\usepackage[usenames,dvipsnames]{color}

\newcommand{\beqn}{\begin{eqnarray}}
\newcommand{\eeqn}{\end{eqnarray}}
\newcommand{\be}{\begin{equation}}
\newcommand{\ee}{\end{equation}}
\newcommand{\beq}{\begin{equation}}
\newcommand{\eeq}{\end{equation}}

\def\s1{$s_{\alpha}$}
\def\s2{$s_{\gamma}$}
\def\s3{$s_{\delta}$}
\def\c1{$c_{\alpha}$}
\def\c2{$c_{\gamma}$}
\def\c3{$c_{\delta}$}

\def\br{\left(\begin{array}{c}}
\def\er{\end{array}\right)}

\begin{document}
\baselineskip 18pt

\thispagestyle{empty}

\vspace{0cm}

\begin{center}
{\bf  \Large
{The Top Quark Electric Dipole Moment in an MSSM  Extension  with  Vector Like  Multiplets } }
\vspace{1.0cm}

{\bf Tarek Ibrahim}\footnote{e-mail: tarek@lepton.neu.edu}$^{,a}$,
{\bf Pran Nath}\footnote{e-mail: nath@lepton.neu.edu}$^{,b}$
\\
\vspace{.5cm}

{\it
$^{a}$  Department of  Physics, Faculty of Science,
University of Alexandria, Alexandria, Egypt\\ 
$^{b}$Department of Physics, Northeastern University,
Boston, Massachusetts 02115, USA \\

}

\end{center}

\vspace{0.2cm}

\begin{center}
{\bf Abstract} \\
\end{center}
\vspace{0cm}
The  electric dipole moment (EDM) of the  top quark is calculated in a model with a vector
like multiplet which mixes with the third generation in an extension of the MSSM. Such mixings allow for new CP violating phases.
Including these new CP phases, the EDM of the  top  in this class of models is computed. The top EDM
arises from loops
involving the exchange of the W, the Z  
 as well as from the exchange involving  the charginos, the neutralinos, the gluino, and the  
 vector like multiplet and their superpartners.
 The analysis of the EDM of the top is more complicated than for the light quarks 
 because the mass of the external fermion,
 in this case the  top quark mass cannot be ignored relative to the masses inside the loops. 
A numerical analysis is presented and it is shown that the top EDM could be close to $10^{-19} ecm$ 
consistent with the current limits on the EDM of the electron, the neutron and on atomic EDMs.
A top EDM of size $10^{-19}ecm$ could be accessible in collider experiments such as the ILC.
\setcounter{footnote}{0}
\clearpage

\section{1. Introduction}
In the Standard Model the EDM of the top quark is rather small
 typically less than $10^{-30}$ ecm\cite{Hoogeveen:1990cb,Soni:1992tn}
 \footnote{The analysis of \cite{Hoogeveen:1990cb} gives  the EDM of the electron
 while the EDM of the top is obtained by scaling the electron EDM  as pointed out by Soni and Xu  in \cite{Soni:1992tn}.}
 In this work we carry out an analysis similar to that 
of \cite{Ibrahim:2010va} (see also  \cite{Ibrahim:2008gg})
 for the EDM of the top quark
arising from the mixing of the third generation with  a vector like generation (For a recent review
on CP violation and on EDMs see \cite{Ibrahim:2007fb}). 
Thus vector  like combinations  are predicted in many unified  models of particle  interactions 
 \cite{Georgi:1979md,Senjanovic:1984rw}
and their implications have been explored in many  works
\cite{Barger:2006fm,Lavoura:1992qd,Maekawa:1995ha,Morrissey:2003sc,Choudhury:2001hs,Liu:2009cc,Babu:2008ge,Martin:2009bg,Graham:2009gy,Arnold:2010vs}.
Such vector like combinations
could lie in the TeV region and  be consistent with the current precision electroweak data.
In this work we allow for the possibility that there could be  a small mixing of these vector like 
combinations with the sequential generations. The implications of such mixings for the neutrino magnetic moment and for the
 anomalous
magnetic moment of the $\tau$ were investigated in\cite{Ibrahim:2008gg} and 
for the EDM of the $\tau$ and $\nu_{\tau}$ in \cite{Ibrahim:2010va}.
In this analysis we  focus on the quark sector of the vector like multiplets. 
To simplify the analysis, we will assume that the mixings of the 
vector like multiplets occurs only with the third generation since such mixings are
consistent with the current experimental constraints\cite{Jezabek:1994zv}.
The masses of the vector multiplets could lie in a large mass range, i.e., from the current lower limits
  given by the  LEP experiment and by the Tevatron up to  a masses lying in the 
   several TeV mass range. 
  Mixings with the vector like generations bring in new phases which are not constrained by the
  current experimental limits on the EDM of the  electron, of the neutron and of atomic EDMs.
    Further details of the vector like models can be found in \cite{Ibrahim:2008gg}  and in the
  other references mentioned therein and above. \\
  
  If vector like multiplets exist and mix with the third generation, they can affect the CP phenomena 
  in the third generation because of the new sources of CP violation arising from mixing with the 
  new sector. 
   The top physics is of course a well known  laboratory for the study of CP violating phenomenona
 ** \cite{Kane:1991bg,Schmidt:1992et,Cuypers:1994ih,Frey:1997sg,atwood,Ibrahim:2003ca}.**
   Specifically the sensitivity of these phenomena to the top EDM has been investigated in a variety of theoretical models
  \cite{Soni:1992tn,as2,Bartl:1997wf,Hollik:1998vz,NovalesSanchez:2009zz,Huang:1994zg}.
  Further, a number of analyses show that in $e^+e^-\to t\bar t$ and in $\gamma \gamma\to t\bar t$ 
  processes the EDM of the top can be measured with great sensitivity 
   \cite{Atwood:1992vj,Poulose:1997xk,Choi:1995kp,Frey:1997sg}, i.e., with 
    sensitivity up to $10^{-19}ecm$
   or even better. 
     With this in mind we investigate here the top EDM in an extension of the MSSM with vector like
   multiplets.  A mixing of the vector like multiplet with the sequential generations and specifically
   the third generation bring in new sources of CP violation which contribute to the top EDM
   and can generate a top EDM as large as $10^{-19}ecm$ well within reach of the sensitivity of 
   the collider experiments. \\
   
   The outline of the  rest of the paper is as follows: In Sec.(2) we  give an analysis of the
 EDM of the top allowing for mixing between the vector like combination and the
third generation quarks. 
The EDM of the  top  arises from loops
involving the exchange of the W, of 
the Z 
as well  as from exchanges involving the charginos, the neutralinos, the gluino as
well as exchanges involving the vector like multiplets and their superpartners. 
 We  note that the analysis of the top EDM is more complicated relative to EDM of the 
light quarks and of the light leptons (see e.g.,\cite{Ibrahim:1997gj}) because we cannot
ignore the mass of the external fermion (i.e., of the top quark in this case) compared to the masses that run
inside the loops. 
So the form factors that enter the analysis of the top EDM  are more complicated relative to 
the form factors that enter the EDM of the light quarks, since for the case of the top the
loop integrals are functions of more than just one mass ratio.
A numerical analysis of the size of the EDM of the top is given in Sec.(3). 
In this section we also  display the dependence of the top EMD on the phases  
and mixings.  Conclusions are given in Sec.(4). Deductions of the mass matrices used
in Sec.(2)  are given in the Appendix. 

\section{Electric dipole moment of the top quark}

\begin{figure}[h!]
\centering
   \includegraphics[width = 16cm, height = 4.5 cm]{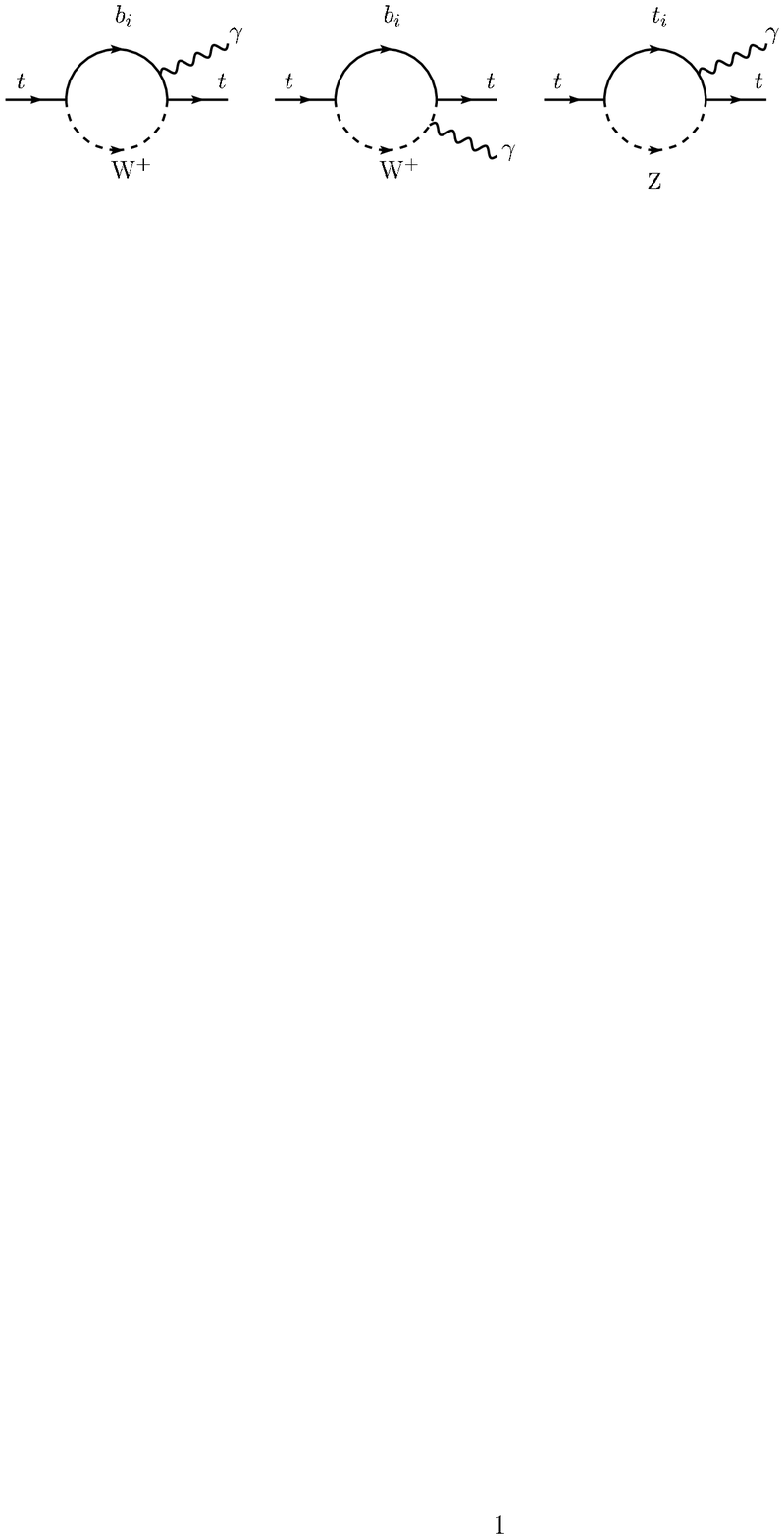}
\caption{One loop contribution to the top electric dipole moment from the 
exchange of the W boson, the Z boson and from the 
exchange of the top and from the exchange of the bottom quarks and their mirrors. 
}
\label{fig1}
\end{figure}

The first and second loops of Fig.(\ref{fig1}) produce EDM of the
 top quark through the interaction of the W boson with
 the top and bottom quarks and the vector multiplet quarks. The relevant part of Lagrangian that 
 generates this contribution is given by 
\beqn
{\cal{L}}_{CC}=-\frac{g}{\sqrt 2} W^{+}_{\mu}
\sum_{i}\sum_{j} \bar t_{j} \gamma^{\mu} 
[D^{t *}_{L1j} D^{b }_{L1i} P_L+ 
 D^{t *}_{R2j} D^{b}_{R2i} P_R] 
 b_{i} +H.c. 
\label{LR}
\eeqn
where $i,j$ run over the set of quarks and mirror quarks including those from the third 
generation and from the vector multiplet,  
 $t_1$ is the physical top quark, and
$D^{t, b}_{L,R}$ are the diagonalizing  matrices defined in the 
 Appendix. These matrices contain phases, and these phases
generate the EDM of the top quark.
Using the above interaction, we get from first and second loops of Fig.(\ref{fig1}), the contribution
\beqn
d^{1(first)}_{t}(W)=-\frac{1}{48\pi^2M^2_W}\sum_{i}m_{b_i} 
Im(\Gamma^{tb}_i)I_1(\frac{m^2_{b_i}}{M^2_W},\frac{m^2_{t_1}}{M^2_W}),\nonumber\\
d^{1(second)}_{t}(W)=-\frac{1}{16\pi^2M^2_W}\sum_{i}m_{b_i} Im(\Gamma^{tb}_i)I_2(\frac{m^2_{b_i}}{M^2_W},\frac{m^2_{t_1}}{M^2_W}).
\label{EDM1}
\eeqn 
Here $\Gamma^{tb}_i$ is given 
\beq
\Gamma^{tb}_i=\frac{g^2}{2}D^{t *}_{L11} D^{b }_{L1i} D^{t}_{R21} D^{b *}_{R2i},
\eeq
and $I_{1,2}(r_1,r_2)$ are given by
\beqn
I_1(r_1,r_2)=\int_0^1 dx \frac{(4+r_1-r_2)x-4x^2}{1+(r_1-r_2-1)x+r_2x^2},\nonumber\\
I_2(r_1,r_2)=\int_0^1 dx \frac{3-5x+(2+r_1-2r_2)x^2+(3+r_2)x^3}{1+(r_1-r_2-1)x+r_2x^2}.
\eeqn
While our analysis is quite general  we will limit ourselves for simplicity 
to the case where
there is mixing between the third generation and the mirror part of the vector multiplet.
The inclusion of the non-mirror part is essentially trivial as it corresponds to an extension of the CKM
matrix from a $3\times 3$ to a $4\times 4$ matrix in the standard  model sector and
similar straightforward extensions in the supersymmetric sector. 
In the rest of the analysis we will focus just
on the mixings with the mirrors which is rather non-trivial.
In fact we work  out, we believe, for the first time the interactions of the quarks-mirrors with the 
charginos, neutralinos and gluinos  which are then utilized in the analysis of Figs.(2).\\

\begin{figure}[h!]
\centering
   \includegraphics[width = 16cm, height = 4 cm]{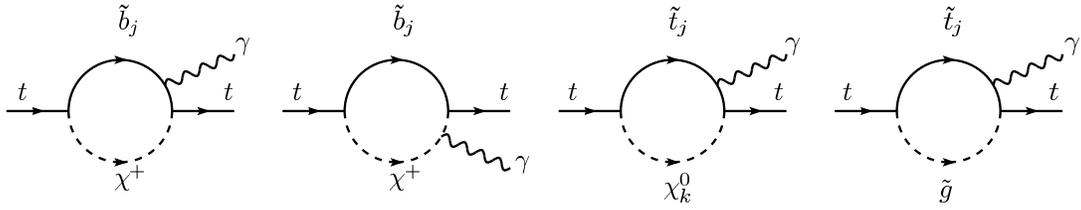}
\caption{One loop contribution to the top electric dipole moment from the 
exchange of the charginos, the neutralinos,  the gluino and from the 
exchange of the stops and from the exchange of the sbottoms  and their mirrors. }
\label{fig2}
\end{figure}

Next we consider the third loop of Fig.(\ref{fig1}) which produces the 
EDM of the top quark through the interaction with the Z 
boson.  The relevant part of Lagrangian that 
 generates this contribution is given by 
\beqn
{\cal{L}}_{NC}=-Z_{\mu}
\sum_{i=1}^2\sum_{j=1}^2\bar t_{j} \gamma^{\mu} 
[S_{Lji} P_L+ 
 S_{Rji} P_R] 
 t_{i}, 
\eeqn
where
\beqn
S_{Lji}=-\frac{g}{6\cos\theta_W}[-3D^{t*}_{L1j}D^t_{L1i}+4\sin^2\theta_W(D^{t*}_{L1j}D^t_{L1i}+D^{t*}_{L2j}D^t_{L2i})],
\nonumber\\
S_{Rji}=-\frac{g}{6\cos\theta_W}[-3D^{t*}_{R2j}D^t_{R2i}+4\sin^2\theta_W(D^{t*}_{R1j}D^t_{R1i}+D^{t*}_{R2j}D^t_{R2i})].
\eeqn
Using the above interaction, we get from third loop of  Fig.(\ref{fig1}), the contribution
\beqn
d^{1(third)}_{t}(Z)=\frac{1}{24\pi^2M^2_Z}\sum_{i=1}^2m_{t_i} Im(S_{L1i}S^*_{R1i})I_1(\frac{m^2_{t_i}}{M^2_Z},\frac{m^2_{t_1}}{M^2_Z}).
\label{EDM2}
\eeqn 
The first and the second loops of Fig.(\ref{fig2}),
 produce EDM of the top through the interaction with 
the charginos. 
The relevant part of Lagrangian that 
 generates this contribution is given by 
\beqn
-{\cal{L}}_{t-\tilde{b}-\chi^+}=
\sum_{k=1}^2\sum_{i=1}^2\sum_{j=1}^4  
\bar{t}_k[\Gamma_{Lkji} P_L+ 
 \Gamma_{Rkji} P_R] 
\tilde{\chi^+}_i \tilde{b}_j +H.c.
\eeqn
where
\beqn
\Gamma_{Lkji}=-g[V^*_{i2}\kappa_t D^{t*}_{R1k}\tilde{D}^b_{1j}-D^{t*}_{R2k}V^*_{i1}\tilde{D}^b_{4j}+D^{t*}_{R2k}\kappa_B V^*_{i2}
\tilde{D}^b_{2j}],
\nonumber\\
\Gamma_{Rkji}=g[U_{i1} D^{t*}_{L1k}\tilde{D}^b_{1j}-D^{t*}_{L1k}\kappa_b U_{i2}\tilde{D}^b_{3j}-D^{t*}_{L2k}\kappa_T U_{i2}
\tilde{D}^b_{4j}],
\eeqn
where $\tilde{D}^b$ is the diagonalizing matrix of the scalar $4\times4$ mass matrices for the scalar quarks as defined in the Appendix.
These elements contain CP violating phases too and can contribute to the EDM of the top.
The couplings $\kappa_f$ are defined as 
\beqn
(\kappa_T, \kappa_{b})
=\frac{(m_T, m_{b})}{\sqrt{2} M_W \cos\beta},~
(\kappa_{B}, \kappa_{t})    =\frac{(m_B, m_{t})}{\sqrt{2} M_W \sin\beta}.
\eeqn
Here  $U$ and $V$ are the matrices  that  diagonalize the chargino mass matrix $M_C$ 
  so that 
\beq
U^* M_C V^{-1}= diag (m_{\tilde{\chi_1}}^+,m_{\tilde{\chi_2}}^+).
\eeq
Using the above interaction, we get from the first and the second loops of Fig.(\ref{fig2})
 the contributions
\beqn
d^{2(first)}_{t}(\chi^+)=-\frac{1}{48\pi^2}\sum_{i=1}^2\sum_{j=1}^4 \frac{m_{\chi^+_i}}{m^2_{\tilde{b_j}}} Im(\Gamma_{L1ji}\Gamma^*_{R1ji})I_3(\frac{m^2_{\chi^+_i}}{m^2_{\tilde{b_j}}},\frac{m^2_{t_1}}{m^2_{\tilde{b_j}}}),
\nonumber\\
d^{2(second)}_{t}(\chi^+)=-\frac{1}{16\pi^2}\sum_{i=1}^2\sum_{j=1}^4 \frac{m_{\chi^+_i}}{m^2_{\tilde{b_j}}} Im(\Gamma_{L1ji}\Gamma^*_{R1ji})I_4(\frac{m^2_{\chi^+_i}}{m^2_{\tilde{b_j}}},\frac{m^2_{t_1}}{m^2_{\tilde{b_j}}}),
\eeqn
where $I_{3,4}(r_1,r_2)$ are given by
\beqn
I_3(r_1,r_2)=\int_0^1 dx \frac{x-x^2}{1+(r_1-r_2-1)x+r_2x^2},\nonumber\\
I_4(r_1,r_2)=\int_0^1 dx \frac{x^2}{1+(r_1-r_2-1)x+r_2x^2}.
\eeqn
We note that the limits of $I_3(r_1,r_2)$ and $I_4(r_1,r_2)$ for $r_2\sim 0$ are the well known form factors $B(r_1)$ and $-A(r_1)$
in the case of light leptons and quarks \cite{Ibrahim:1997gj}.\\

The third loop of Fig.(\ref{fig2}) produces EDM of the top  through the interaction of the
 neutralinos.  The relevant part of 
Lagrangian that generates this contribution is given by 
\beqn
-{\cal{L}}_{t-\tilde{t}-\chi^0}=
\sum_{k=1}^4\sum_{i=1}^4\sum_{j=1}^2  
\bar{t}_j[C_{Ljki} P_L+ 
C_{Rjki} P_R] 
\tilde{\chi^0}_i \tilde{t}_k +H.c.,
\eeqn
where
\beqn
C_{Ljki}=\sqrt{2}[\alpha_{ti}D^{t *}_{R1j} \tilde{D}^t_{1k}-\gamma_{ti}D^{t *}_{R1j} \tilde{D}^t_{3k}
+\beta_{Ti}D^{t *}_{R2j} \tilde{D}^t_{4k}-\delta_{Ti}D^{t *}_{R2j} \tilde{D}^t_{2k}],
\nonumber\\
C_{Rjki}=\sqrt{2}[\beta_{ti}D^{t *}_{L1j} \tilde{D}^t_{1k}-\delta_{ti}D^{t *}_{L1j} \tilde{D}^t_{3k}
+\alpha_{Ti}D^{t *}_{L2j} \tilde{D}^t_{4k}-\gamma_{Ti}D^{t *}_{L2j} \tilde{D}^t_{2k}].
\eeqn
The matrix  $\tilde{D}^t$ is the diagonalizing matrix of the $4\times 4$ stop mixed with scalar mirrors mass$^2$ matrix as shown in the Appendix.
The couplings that enter the above equations are given by
\beqn\label{alphabk}
\alpha_{t j} =\frac{g m_{t} X_{4j}}{2m_W\sin\beta},~~
\beta_{t j}=\frac{2}{3}eX_{1j}^{'*} +\frac{g}{\cos\theta_W} X_{2j}^{'*}
(\frac{1}{2}-\frac{2}{3}\sin^2\theta_W),\nonumber\\
\gamma_{tj}=\frac{2}{3}e X_{1j}^{'}-\frac{2}{3}\frac{g\sin^2\theta_W}{\cos\theta_W}
X_{2j}^{'},
~~ \delta_{t j}=-\frac{g m_{t} X_{4j}^*}{2m_W \sin\beta}.
\eeqn
Here 
\beqn
\alpha_{T j} =\frac{g m_{T} X^*_{3j}}{2m_W\cos\beta},~~
\beta_{T j}=-\frac{2}{3}eX_{1j}^{'} +\frac{g}{\cos\theta_W} X_{2j}^{'}
(-\frac{1}{2}+\frac{2}{3}\sin^2\theta_W),\nonumber\\
\gamma_{Tj}=-\frac{2}{3}e X_{1j}^{'*}+\frac{2}{3}\frac{g\sin^2\theta_W}{\cos\theta_W}
X_{2j}^{'*},
~~ \delta_{T j}=-\frac{g m_{T} X_{3j}}{2m_W \cos\beta},
\eeqn
where
\beqn
X'_{1j}= (X_{1j}\cos\theta_W + X_{2j} \sin\theta_W), 
~X'_{2j}=  (-X_{1j}\sin\theta_W + X_{2j} \cos\theta_W), 
\eeqn
and where the matrix $X$ diagonlizes the neutralino mass matrix so that
\beq
X^T M_{\tilde{\chi}^0} X=diag(m_{{\chi^0}_1}, m_{{\chi^0}_2}, m_{{\chi^0}_3}, m_{{\chi^0}_4}).
\eeq
Using the above interaction, we get from the third loop of Fig.(\ref{fig2})
 the neutralino contributions to the top  EDM to be
\beqn
d^{2(third)}_{t}(\chi^0)=\frac{1}{24\pi^2}\sum_{i=1}^4\sum_{k=1}^4 \frac{m_{\chi^0_i}}{m^2_{\tilde{t_k}}} Im(C_{L1ki} C^*_{R1ki})I_3(\frac{m^2_{\chi^0_i}}{m^2_{\tilde{t_k}}},\frac{m^2_{t_1}}{m^2_{\tilde{t_k}}}).
\eeqn

Finally the gluino contribution to the electric dipole moment of the top  comes from the fourth loop of 
Fig.(\ref{fig2}). 
 The relevant part of Lagrangian that  generates this contribution is given by  
\beqn
-{\cal{L}}_{t\tilde{t} \tilde{g}}=\sqrt{2} g_s
\sum_{a=1}^8\sum_{j,k=1}^3\sum_{n=1}^2\sum_{m=1}^4 T^a_{jk} \bar t^j_{n}  
[K_{L_{nm}} P_L+K_{R_{nm}}P_R] \tilde{g}_a \tilde{t}^k_m +H.c. 
\eeqn
where
\beqn
K_{L_{nm}}=e^{-i\xi_3 /2}[D^{t*}_{R_{2n}} \tilde{D}^{t}_{4m}-D^{t*}_{R_{1n}}\tilde{D}^{t}_{3m}],\nonumber\\
K_{R_{nm}}=e^{i\xi_3 /2}[D^{t*}_{L_{1n}} \tilde{D}^{t}_{1m}-D^{t*}_{L_{2n}}\tilde{D}^{t}_{2m}],
\eeqn
where $\xi_3$ is the phase of the 
 gluino mass.
The above Lagrangian gives a contribution  
\beqn
d^{2(fourth)}_{t}(\tilde{g})=\frac{g^2_s}{9\pi^2}\sum_{j=1}^4 \frac{m_{\tilde{g}}}{m^2_{\tilde{t_j}}} 
Im(K_{L_{1j}} K^*_{R_{1j}})I_3(\frac{m^2_{\tilde{g}}}{m^2_{\tilde{t_j}}},\frac{m^2_{t_1}}{m^2_{\tilde{t_j}}}).
\eeqn

\section{Numerical  Analysis}
The mixing matrices between the quarks and the mirrors are diagonalized using bi-unitary matrices (see the Appendix). 
So we parametrize the mixing between $t$ and $T$  by the angles $\theta_L$, $\theta_R$, $\chi_L$ and $\chi_R$, 
and the mixing between $b$ and $B$ by the angle $\phi_L$, $\phi_R$, $\xi_L$ and $\xi_R$
where 
\beqn
D^{t}_L=
 {\left(
\begin{array}{cc}
\cos\theta_L & -\sin\theta_L e^{-i\chi_L} \cr
             \sin\theta_L  e^{i\chi_L}& \cos\theta_L
\end{array}\right)},
~D^{b}_L=
 {\left(
\begin{array}{cc}
\cos\phi_L & -\sin\phi_L e^{-i\xi_L} \cr
             \sin\phi_L  e^{i\xi_L}& \cos\phi_L
\end{array}\right)},
\eeqn
and $D^{t}_R$ and $D^{b}_R$ can be gotten from $D^{t}_L$ and $D^{b}_L$
by the following substitution:  
$D^{t}_L\to  D^{t}_R, \theta_L\to \theta_R,  \chi_L\to \chi_R$,    and 
$D^{b}_L \to D^{b}_R, \phi_L\to \phi_R, \xi_L\to \xi_R$.
We note that  the phases $\chi_{L,R}$ arise from the couplings $h_3$ and $h_5$ 
while the phases $\xi_{L,R}$ arise from the couplings $h_4$ and $h_3$ through 
the relations
\beqn
\chi_R=arg (-m_{t} h_3+m_T h^*_5),
~\chi_L=arg (m_{t}h^*_5-m_T h_3),\nonumber\\
~\xi_R=arg (m_{b}h_3+m_B h^*_4),
~\xi_L=arg (m_{b}h^*_4+m_B h_3).
\eeqn
For the case of top and bottom masses arising from hermitian matrices, i.e., when
 $h_5=-h^*_3$ and $h_4=h^*_3$ we have
$\theta_L=\theta_R$, $\phi_L=\phi_R$, $\chi_L=\chi_R=\chi$ and $\xi_L=\xi_R=\xi$.  Further,   here we have
the relation $\xi=\chi+\pi$ and thus the W-exchange and Z-exchange terms in the EDM for the top  
vanish.  However, more generally  the top and the bottom mass matrices are not hermitian and  they generate 
non-vanishing  contributions to the EDMs.
Thus the input parameters for this sector of the parameter space are
$m_{t1},m_T, h_3, h_5, m_{b1}, m_B, h_4$
with $h_3$, $h_4$ and $h_5$ being  complex masses with the corresponding 
CP violating phases $\chi_3$, $\chi_4$ and 
$\chi_5$.
For the sbottom and stop mass$^2$ matrices we need the extra input parameters of the susy breaking sector,
$
\tilde{M}_q, \tilde{M}_B,\tilde{M}_{b},\tilde{M}_{Q},\tilde{M}_{t},\tilde{M}_T,
A_{b}, A_T, A_{t},A_{B}, \mu, \tan\beta.
$
The chargino, neutralino and gluino sectors need the extra parameters
$\tilde{m}_1, \tilde{m}_2$ and $m_{\tilde{g}}$.
We will assume that the only parameters that have phases in the above set are 
$A_T$, $A_B$, $A_{t}$ and $A_{b}$ with the corresponding phases given by 
 $\alpha_T$, $\alpha_B$, $\alpha_{t}$ and $\alpha_{b}$.\\
 
\begin{figure}[h!]
\centering
   \includegraphics[width = 9cm, height = 5 cm,angle=270]{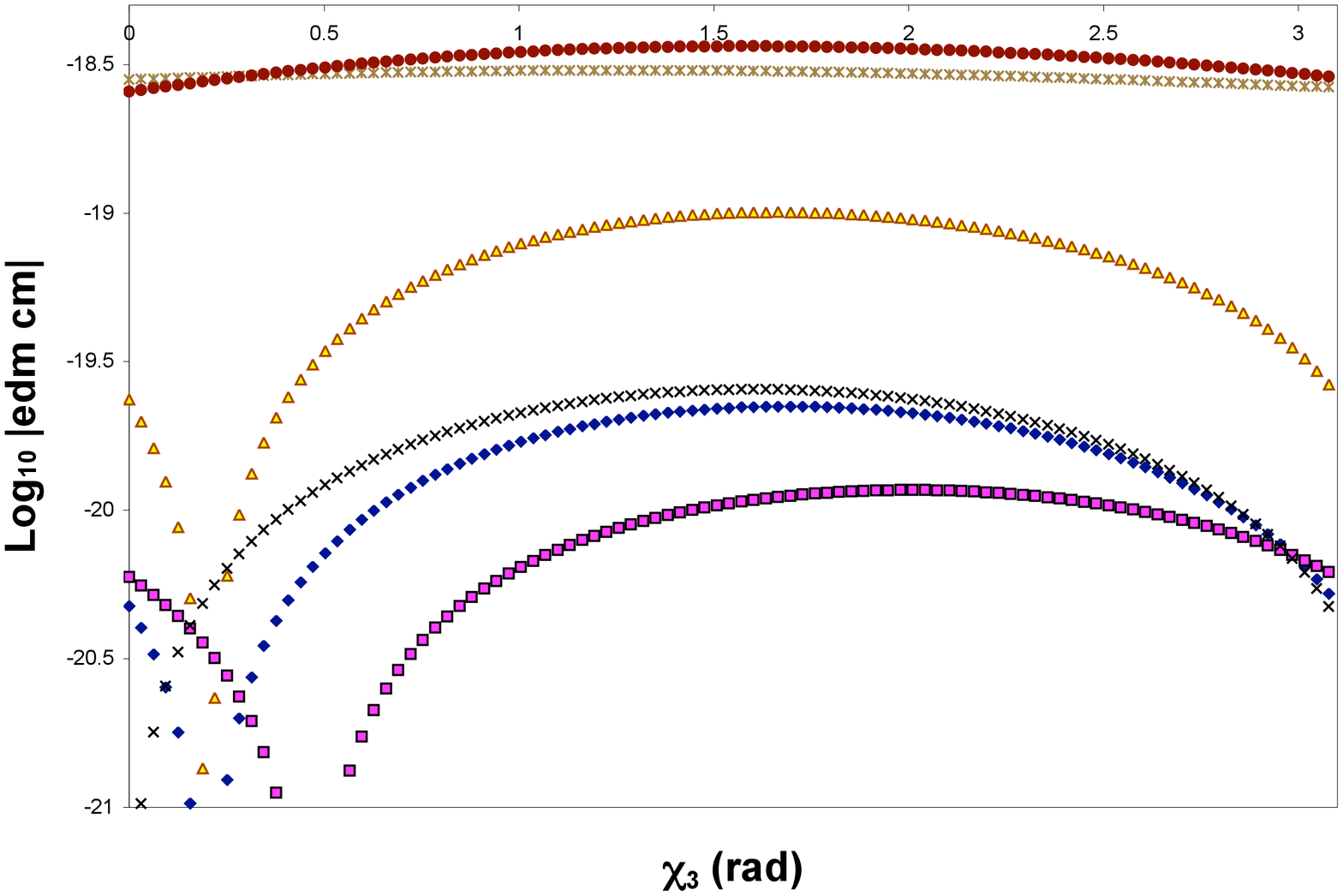}
   \hspace{2cm}
      \includegraphics[width = 9cm, height = 5 cm,angle=270]{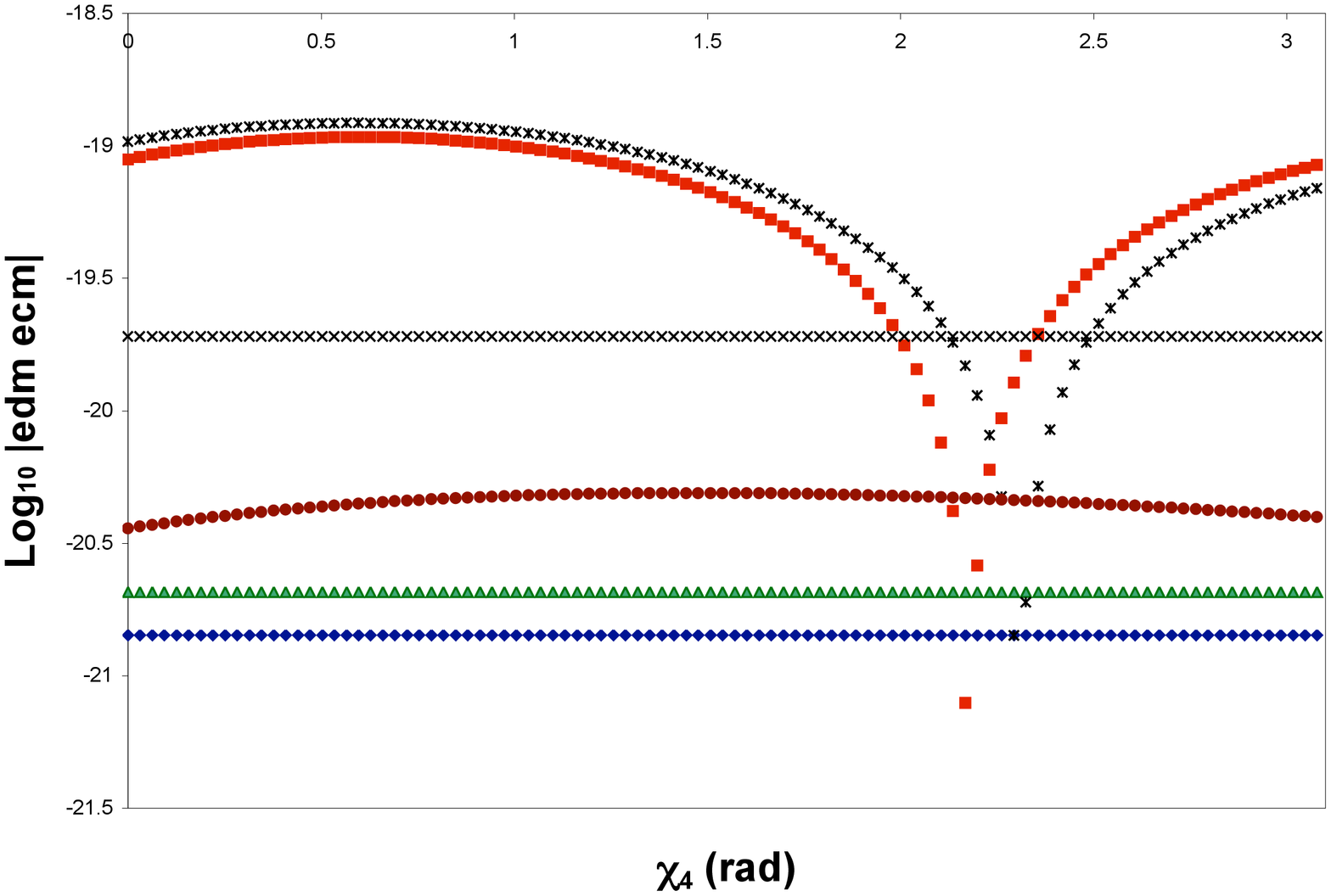}
      \vspace{-2cm}
      \caption{ 
\scriptsize
Left:  An exhibition of the dependence of $d_{t}$ on $\chi_3$ when $\tan\beta=10$, 
$m_T=300$, $|h_3|=$80, $|h_4|=$70, $m_B=$150,
$|h_5|=$90, $m_0=$100, $|A_0|=$150, $\tilde{m}_1=50$, $\tilde{m}_2=100$, 
$\mu=150$, $\tilde{m}_g=400$, $\chi_4=$0.7, $\chi_5=-$0.6, $\alpha_T=$0.7,   and $\alpha_B=$0.1.
( The six curves correspond to the contributions from
 the neutralino, W, Z, chargino, total EDM and gluino. They are shown in ascending order at $\chi_3=0$). 
Here and in subsequent figures all
masses are in GeV and  all angles are in rad. Right:
An exhibition of the dependence of $d_{t}$ on $\chi_4$ when $\tan\beta=20$, 
$m_T=400$, $|h_4|=$60, $m_B=$100, $|h_3|=$70,
$|h_5|=$60, $m_0=$150, $|A_0|=$100, $\tilde{m}_1=50$, $\tilde{m}_2=100$, $\mu=150$, $\tilde{m}_g=400$, $\chi_3=$0.5, $\chi_5=$0.3, $\alpha_T=$0.1 and  $\alpha_B=$0.7( The six curves correspond to the contributions from
 the Z, neutralino, W, gluino, chargino and total EDM. They are shown in ascending order at $\chi_4=0$). 
}
\label{fig3}
\end{figure}

To simplify the analysis further we set some of the  phases to zero, i.e., specifically we set 
$\alpha_{t}=\alpha_{b}=0$.
 With this in mind
the only contributions to the EDM of the top quark arises from mixing terms between the scalars and the
 mirror scalars, between the fermions - and the mirror fermions and finally among the mirror scalars themselves.
Thus in the absence of the mirror part of the lagrangian, the top EDM  vanishes and so we can isolate
the  role of the CP violating phases in this sector and see the size of its contribution.
The $4\times 4$ mass$^2$ matrices of stops and sbottoms are diagonlized numerically.
Thus the CP violating phases that would play a role in this analysis are
\beq
\chi_3, \chi_4, \chi_5, \alpha_T, \alpha_B.
\eeq
To reduce the number of input parameters we assume
$\tilde{M}_a =m_0, ~a=q, B, b, Q, T, t$ and $|A_i|=|A_0|$, $i=T, B, t, b$. 
In  the left panel of  Fig(\ref{fig3}), we  give a numerical analysis of the top EDM  and discuss
its  variation with the phase $\chi_3$. 
 We note that 
 $\chi_3$ enters $D^{t}$, $D^{b}$, $\tilde D^{t}$ and $\tilde D^{b}$ and as a consequences all diagrams
in Fig.(1) and in Fig. (2) that contribute to the top EDM have a $\chi_3$ dependence. \\

Further, the  
 various diagrams that contribute to the top
 EDM may add constructively or destructively as shown in the Z, W, neutralino and chargino contributions. 
In the case of destructive interference, we have large
cancellations reminiscent of the cancellation mechanism for the EDM of the electron and for 
the neutron\cite{incancel,lpr}.
Of course the desirable larger contributions  for the top  EDM occur away
from the cancellation regions.  
In the right panel of  Fig(\ref{fig3}), we study the variation of the different components of $d_t$ as the magnitude 
of the phase $\chi_4$ varies. 
The sparticle masses and couplings in the bottom sector and thus  the top
  EDM arising  from the exchange of the W and the charginos  are
  sensitive to $\chi_4$ and thus only these two  contributions to the top EDM  
  have dependence on  this parameter.\\

\begin{figure}[h!]
\centering
   \includegraphics[width = 9cm, height = 5 cm,angle=270]{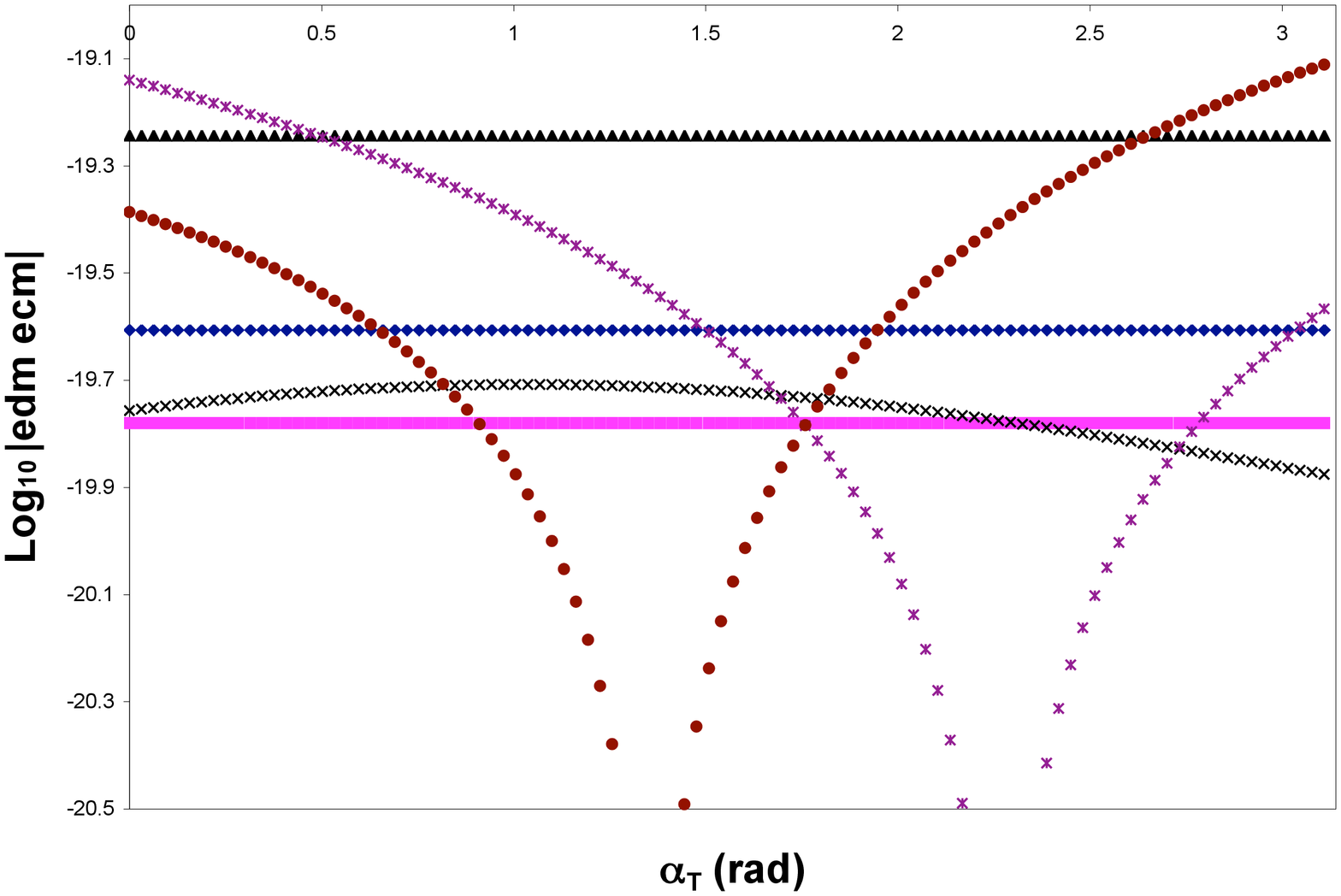}
   \hspace{2cm}
      \includegraphics[width = 9cm, height = 5 cm,angle=270]{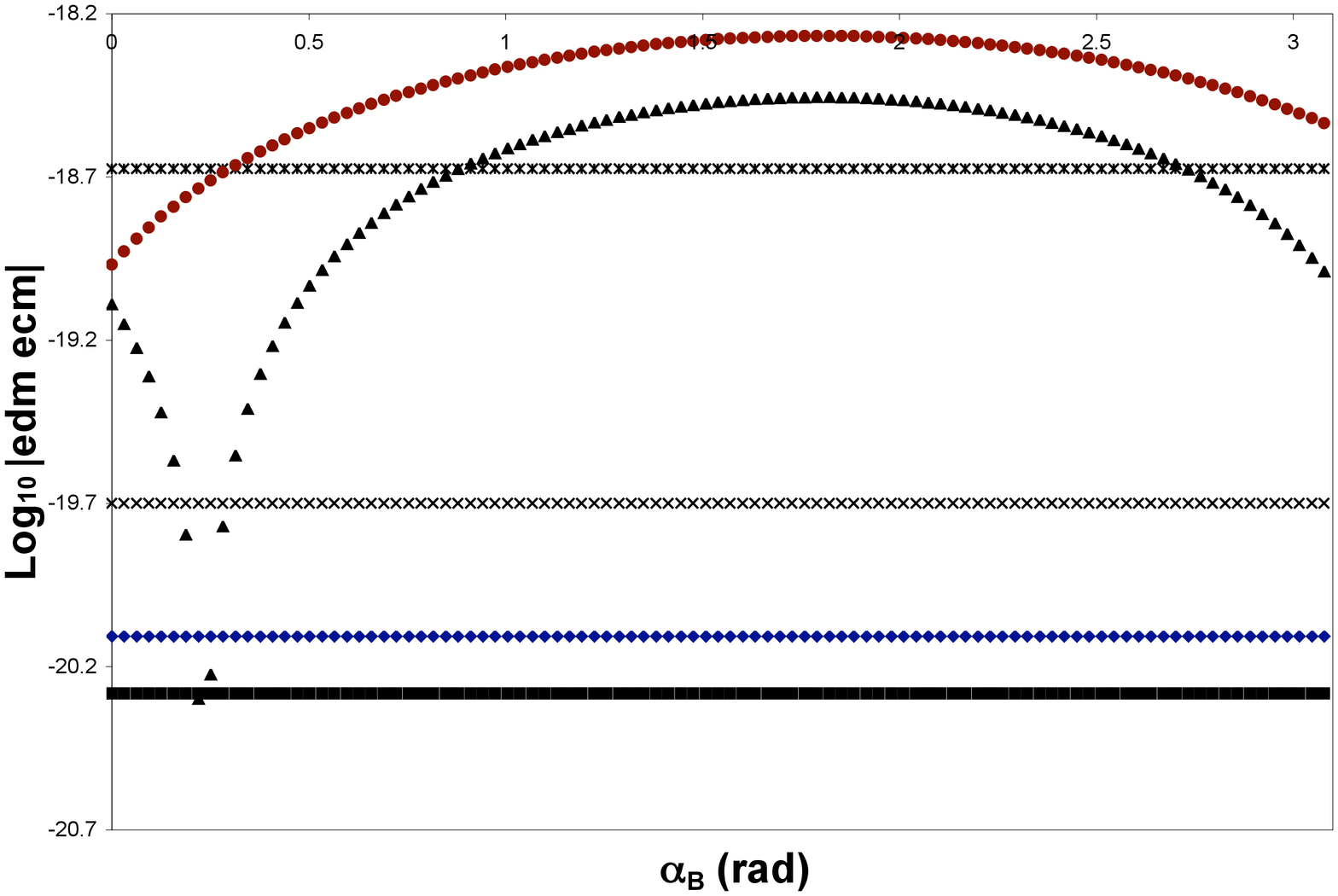}
      \vspace{-2cm}
      \caption{ 
\scriptsize
Left:
An exhibition of the dependence of $d_{t}$ on $\alpha_T$ when $\tan\beta=15$, 
$m_T=350$, $|h_3|=$85, $m_B=$150,
$|h_4|=$75, $|h_5|=$65, $m_0=$200, $|A_0|=$200, $\tilde{m}_1=50$, $\tilde{m}_2=100$, $\mu=150$, $\tilde{m}_g=400$, $\chi_3=$0.5 $\chi_4=$0.5, $\chi_5=-$0.6, and  $\alpha_B=-$0.4( The six curves correspond to the 
 the Z, neutralino, W, total EDM, chargino and gluino contributions. They are shown in ascending order at $\alpha_T=0$). 
Right: 
An exhibition of the dependence of $d_{t}$ on $\alpha_B$ when $\tan\beta=5$, 
$m_T=280$, $|h_3|=$90, $m_B=$200,
$|h_4|=$80, $|h_5|=$70, $m_0=$220, $|A_0|=$250, $\tilde{m}_1=50$, $\tilde{m}_2=100$, $\mu=150$, $\tilde{m}_g=400$, $\chi_3=$0.5 $\chi_4=$0.4, $\chi_5=-$0.5, and  $\alpha_T=$1.2( The six curves correspond to the 
 the Z, W, neutralino, chargino, total EDM and gluino contributions. They are shown in ascending order at $\alpha_B=0$). 
 }
\label{fig4}
\end{figure}

\begin{figure}[h!]
 \vspace{-3cm}
 \centering
   \includegraphics[width = 11cm, height = 7 cm,angle=270]{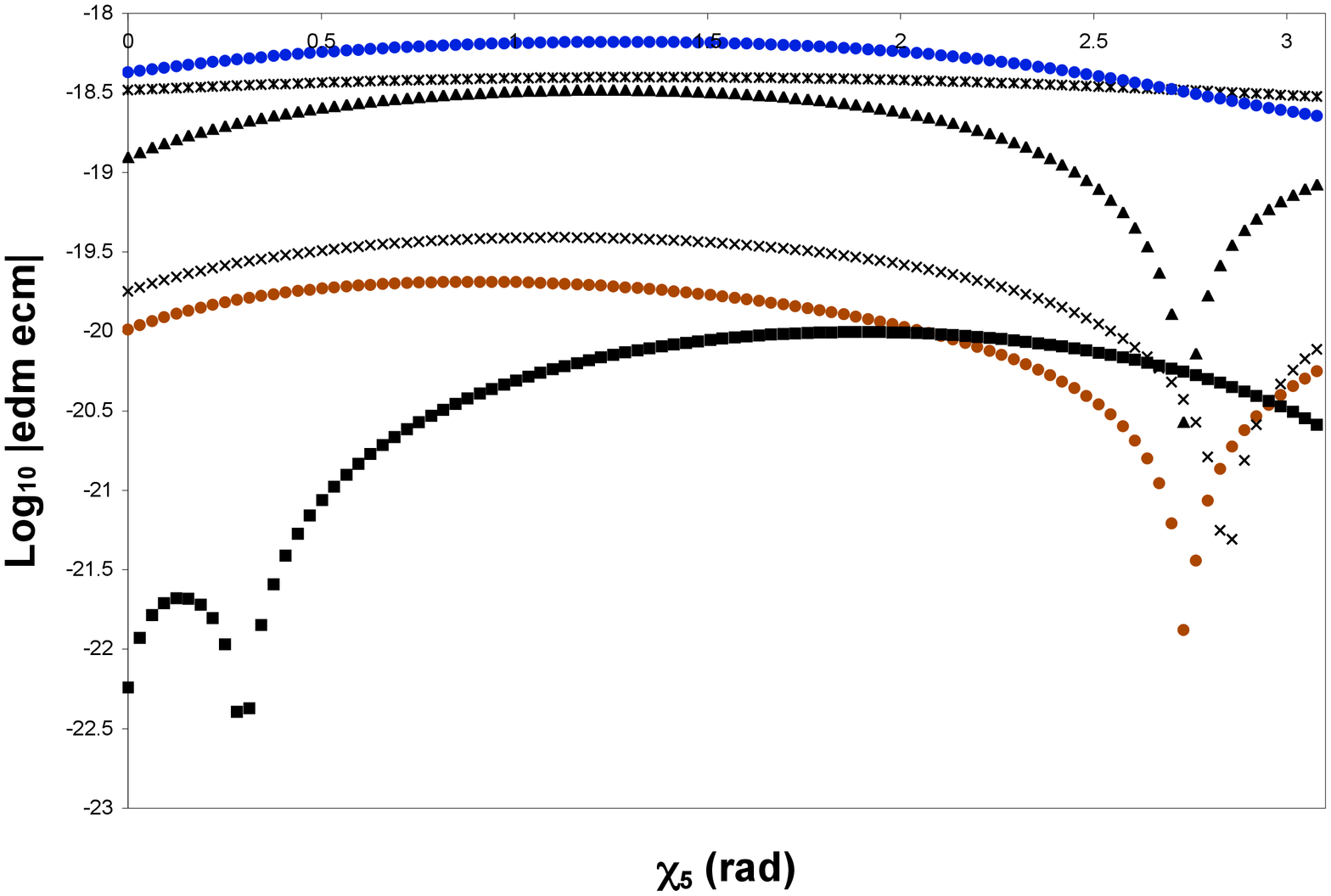}
   \vspace{-3cm}
\caption{ 
\scriptsize
 An exhibition of the dependence of $d_{t}$ on $\chi_5$ when $\tan\beta=7$, 
$m_T=500$, $|h_3|=$60, $m_B=$120,
$|h_4|=$70, $|h_5|=$80, $m_0=$100, $|A_0|=$200, $\tilde{m}_1=50$, $\tilde{m}_2=100$, $\mu=150$, $\tilde{m}_g=400$, $\chi_3=$0.3 $\chi_4=$0.5, $\alpha_B=$.2 and  $\alpha_T=$.7( The six curves correspond to the 
 the Z, W, neutralino, chargino, gluino and total EDM. They are shown in ascending order at $\chi_5=0$). 
}
\label{fig5}
\end{figure}

The left panel of Fig(\ref{fig4}) exhibits   the variation of the different components of  $d_t$ on the  phase $\alpha_T$.
We observe that the  components that vary with this phase are the neutralino and the gluino contributions while
the W, Z and chargino contributions have no dependence on  this phase. The reason for the above is 
that  $\alpha_T$  enters
the  scalar top mass$^2$ matrix and the EDM arising from  W, Z and chargino exchanges are 
independent of $\tilde{D}^t$. 
However, 
 the neutralino and the gluino contributions are affected by it.
It is clear that we see here too the cancellation mechanism working since the components are close to each other
with different signs, so we have the possibility of a destructive cancellation.
In the right panel of Fig(\ref{fig4}), we study the variation of the different components of  $d_t$ as the phase $\alpha_B$ changes.
We note that the only component that varies with this phase is the chargino component. This is expected since
$\alpha_B$ 
 enters the scalar bottom mass$^2$ matrix and  the chargino
contribution to the EDM is  controlled by   $\tilde{D}^b$ which depends on $\alpha_B$ while  the other 
contributions are independent of this phase.
In  Fig(\ref{fig5}) we study the variation of the different components of  $d_t$ as the phase $\chi_5$ changes.
This phase enters  the top quark mass matrix and the scalar top mass$^2$ matrix and consequently 
 the matrices $D_{L,R}^t$ and $\tilde{D}^t$. Thus the contributions to the EDM of the top arising from the
 W, Z, neutralino, chargino and gluino exchanges  all have a dependence on $\chi_5$ as exhibited in
Fig(\ref{fig5}). \\

As mentioned in the introduction the top EDM can be explored in the $e^+e^-\to t\bar t$ and 
$\gamma\gamma\to t\bar t$ processes
  \cite{Atwood:1992vj,Poulose:1997xk,Choi:1995kp,Frey:1997sg}.
  Specifically, it is demonstrated that at a linear $e^+e^-$ collider (such as the ILC), 
  one can explore the top EDM at the level of   $10^{-19}-10^{-20}ecm$
  \footnote{The analysis of\cite{Frey:1997sg} indicates that an $e^+e^-$ collider 
  at $\sqrt{s}=500$ GeV with $10$ fb$^{-1}$ of integrated  luminosity will be sensitive to the $t-Z$
  electric dipole moment up to $8\times 10^{-20} ecm$.}.
   Thus the top EDM predicted in the model with extra vector multiplets falls within the 
   realm of exploration in future collider experiments.
   In Table (1) we give a sample exhibition of
the size and the sign of every component to the top quark EDM.
The analysis of Table (1) and of Fig.(\ref{fig3}), Fig.(\ref{fig4}) and Fig.(\ref{fig5}) 
show that a top EDM as large $10^{-19}$ecm can be gotten which falls within reach of the experiments
at the ILC. It would be interesting to  explore also the sensitivity of the LHC experiments to 
the  top EDM. 
Finally we note that models of the type discussed here  can produce interesting signatures at the LHC some of which are discussed in\cite{Ibrahim:2008gg} and \cite{Nath:2010zj}.\\

\begin{center} \begin{tabular}{|c|c|c|c|c|c|}
\multicolumn{6}{c} {Table~1:  $W^+$, $Z$, $\chi^{+}$, $\chi^0$, $\tilde g$ exchange contributions to 
the top EDM $d_t$. } \\
\hline
$\chi_3 (rad)$ & $d_t(W) e.cm$  &$d_t (Z) e.cm$ & $d_t({\chi^+})e.cm$ &
$d_t({\chi^0}) e.cm$ & $d_t(\tilde{g}) e.cm$  \\
\hline
\hline
$0.0$     &  $4.73\times 10^{-21}$  &  $-5.94\times 10^{-21}$     &   $-7.04\times 10^{-20} $
&    $-2.85\times 10^{-22}$ & $2.98\times 10^{-19}$
\\
 \hline
$0.5$     &  $-7.11\times 10^{-21}$  &  $4.84\times 10^{-22}$     &   $1.02\times 10^{-19} $
&    $-1.21\times 10^{-20}$ & $3.08\times 10^{-19}$
\\
\hline
$1.0$     &  $-1.69\times 10^{-20}$  &  $6.38\times 10^{-21}$     &   $2.36\times 10^{-19} $
&    $-2.12\times 10^{-20}$ & $3.11\times 10^{-19}$
\\
\hline
$1.5$     &  $-2.19\times 10^{-20}$  &  $1.03\times 10^{-20}$     &   $2.98\times 10^{-19} $
&    $-2.53\times 10^{-20}$ & $3.07\times 10^{-19}$
\\
\hline
$2.0$     &  $-2.13\times 10^{-20}$  &  $1.16\times 10^{-20}$     &   $2.87\times 10^{-19} $
&    $-2.37\times 10^{-20}$ & $2.96\times 10^{-19}$
\\
\hline
 \hline
\end{tabular}\\~\\
\label{tab:1}
\noindent
\end{center}
Table caption:  A sample illustration of the various 
contributions to the electric dipole moment of the  top quark.  The inputs are:  $\tan\beta=10$, 
$m_T=300$, $|h_3|=$80, $|h_4|=$70, $m_B=$150,
$|h_5|=$90, $m_0=$100, $|A_0|=$150, $\tilde{m}_1=50$, $\tilde{m}_2=100$, 
$\mu=150$, $\tilde{m}_g=400$, $\chi_4=$0.7, $\chi_5=-$0.6, $\alpha_T=$0.7,   and $\alpha_B=$0.1.
All masses are in units of GeV and all angles are in radian.

\section{Conclusion}
As is well known EDMs are probes of new physics beyond the Standard Model.  
In this paper we have given an analysis of the EDM of the top quark in an extended MSSM model
which includes an extra vector  like multiplet containing quarks and mirror quarks  and their 
superpartners  which can mix with the third generation. 
 A small mixings of this type is not excluded by experiment
 for the third generation. Such mixings bring in 
 new sources of CP violation which do not enter in the analysis of the EDMs of the first two 
generations. Thus these phases are typically unconstrained and  can be large.
The relevent parts of the MSSM Lagrangian interactions involving mirror quarks,  
 charginos, neutralinos and gluinos have been worked out, we believe,   for the first time in this analysis.
 The   EDM of the top in computed using these interactions. The analysis has many one loop diagrams
involving  the exchange of the W, the Z  as shown in Fig.(1),
and also from loops involving the exchange of  the charginos, the neutralinos, the gluino 
as shown in Fig.(2),  each involving also the exchange of vector 
multiplets in the loops.
It is found that the  analysis of the top EDM is more involved in this case as compared to the 
 analysis for the light quarks or for the leptons. This is so  because the mass of the external 
 fermion cannot be ignored relative to the mass of the exchanged particles in this case, 
 as their masses are  comparable.  Because of this, the loop integrals  in this case are  functions of 
 more than just one mass ratio. Finally, we have carried out a numerical analysis of the top EDM 
 in this model. 
 The analysis shows that with large CP phases arising  from the new sector the top EDM can be
 as large as $10^{-19} ecm$ which can be probed in processes such as 
 $e^+e^-\to t\bar t$ and $\gamma \gamma \to t\bar t$ in  collider experiments. 
 It should be interesting to also analyse the sensitivity that the LHC can achieve for the top EDM. 
 Finally, we note  that  the contributions of the chromoelectric dipole moment and of the purely gluonic dimension
 six operator  were not considered in the current work.
 These contributions  entail computations of a new  set of diagrams involving external 
 gluon vertices and require a separate analysis. 
 However, we expect these contributions  to be  of 
 similar  size as the one computed here. \\
 
\noindent
{\em Acknowledgments}:  
This research is  supported in part by  NSF grant PHY-0757959 and by PHY-0704067.
\section{Appendix: Mass matrices  for quarks and squarks and for  their mirrors}
In this Appendix we exhibit 
 the mass matrices for the top quarks, the bottom quarks, the scalar quarks 
and their mirrors  that
enter in the computations of the EDM of the top.  In the deduction of the  
  mass  matrices we need the transformation properties of the quarks and  their mirrors.
Thus under $SU(3)_C\times SU(2)_L\times U(1)_Y$ the quarks transform as 
follows
\beqn
q\equiv 
 \left(\matrix{ t_L\cr 
 b_L}\right)
\sim(3,2,\frac{1}{6}), t^c_L\sim (3^*,1,-\frac{2}{3}), b^c_L\sim (3^*,1,\frac{1}{3}),
\eeqn
while  the mirror quarks transform as 
\beqn
Q^c \equiv 
 \left(\matrix{ B^c_L\cr 
 T^c_L}\right)
\sim(3^*,2,-\frac{1}{6}), T_L\sim (3,1,\frac{2}{3}), B_L\sim (3^*,1, -\frac{1}{3}).
\eeqn
For the Higgs multiplets  we have the MSSM Higgs  doublets with the $SU(3)_C\times SU(2)_L\times U(1)_Y$ 
transformation properties as follows 
\beqn
H_1\equiv 
 \left(\matrix{ H_1^1\cr 
 H_1^2}\right)
\sim(1,2,-\frac{1}{2}), ~H_2\equiv 
 \left(\matrix{ H_2^1\cr 
 H_2^2}\right)
\sim(1,2,\frac{1}{2}).
\eeqn

We assume that the mirrors of the vector like generation escape acquiring mass at the GUT scale and remain
light down to the electroweak scale where the superpotential of the model
for the quark part  may be written  in the form
\beqn
W= \epsilon_{ij}  [h_{1} \hat H_1^{i} \hat q_L ^{j}\hat b^c_L
 +h_{1}' \hat H_2^{j} \hat q_L ^{i} \hat t^c_L
+h_{2} \hat H_1^{i} \hat Q^c{^{j}}\hat T_{L}
 +h_{2}' \hat H_2^{j} \hat Q^c{^{i}} \hat B_L]\nonumber\\
+ h_{3} \epsilon_{ij}  \hat Q^c{^{i}}\hat q_L^{j}
 + h_{4} \hat b^c_L \hat  B_L  +  h_{5} \hat t^c_L \hat T_{L}.
\label{superpotential}
\eeqn
Mixings of the above type can arise 
 via non-renormalizable interactions \cite{Senjanovic:1984rw}.
To get the mass matrices of the quarks and of the mirror quarks we 
replace the superfields in the superpotential by their component scalar
fields. The relevant parts in the superpotential that produce the quark and
mirror quark mass matrices are
\beqn
W=h_1 H_1^1 \tilde{b}_L \tilde{b}_R^* +h_1' H_2^2 \tilde{t}_{L} \tilde{t}^*_{R}+
h_2 H_1^1 \tilde{T}_R^* \tilde{T}_L+h_2' H_2^2 \tilde{B}_R^* \tilde{B}_{L}\nonumber\\
+h_3 \tilde{B}^*_{R} \tilde{b}_L -h_3 \tilde{T}_R^* \tilde{t}_{L}+ h_4 \tilde{b}_R^* \tilde{B}_{L}
+h_5 \tilde{t}^*_{R} \tilde{T}_L.
\eeqn
The mass terms for the quarks and their mirrors arise from the part of the lagrangian
\beq
{\cal{L}}=-\frac{1}{2}\frac{\partial ^2 W}{\partial{A_i}\partial{A_j}}\psi_ i \psi_ j+H.c.
\eeq
where $\psi$ and $A$ stand for generic two-component fermion and scalar fields.
After spontaneous breaking of the electroweak symmetry, ($<H_1^1>=v_1 $ and $<H_2^2>=v_2$),
we have the following set of mass terms written in the 4-spinor notation for the fermionic sector 
\beqn
-{\cal L}_m = \br\bar b_R ~ \bar B_{R} \er
 \br
  h_1 v_1 ~~ h_4\\
 h_3 ~~ h_2' v_2\er
 \br b_L\\
 B_{L}\er
  + \br\bar t_R ~~ \bar T_R\er
 \br h'_1 v_2~~ h_5\\
 -h_3 ~~ h_2 v_1\er \br t_L\\
 T_L\er  + H.c.\nonumber
\eeqn 
Here 
the mass matrices are not  Hermitian and one needs
to use bi-unitary transformations to diagonalize them. Thus we write the linear transformations
\beqn
 \br b_R\\ 
 B_{R}\er=D^{b}_R \br b_{1_R}\\
 b_{2_R} \er,
~\br b_L\\
 B_{L}\er=D^{b}_L \br b_{1_L}\\
 b_{2_L}\er,
\eeqn
such that
\beq
D^{b\dagger}_R \br h_1 v_1 ~~ h_4\\
 h_3 ~~ h_2' v_2\er D^{b}_L=diag(m_{b_1},m_{b_2}),
\label{put1}
\eeq
and the same holds for the top mass matrix so that 
\beq
D^{t \dagger}_R \br h'_1 v_2 ~~ h_5\\
 -h_3 ~~ h_2v_1\er D^{t}_L=diag(m_{t_1},m_{t_2}).
\label{put2}
\eeq

Here $b_1, b_2$ are the mass eigenstates and we identify the bottom quark 
with the eigenstate 1, i.e.,  $b=b_1$, and identify $b_2$ with a heavy 
mirror eigenstate  with a mass in the hundreds  of GeV. Similarly 
$t_1, t_2$ are the mass eigenstates for the top quarks, 
where we identify $t_1$ with the physical top quark with the lighter mass,  and $t_2$ with the 
heavier mass eigenstate.
By multiplying Eq.(\ref{put1}) by $D^{b \dagger}_L$ from the right and by
$D^{b}_R$ from the left and by multiplying Eq.(\ref{put2}) by $D^{t \dagger}_L$
from the right and by $D^{t}_R$ from the left, one can equate the values of the parameter
$h_3$ in both equations and we can get the following relation
between the diagonalizing matrices $D^{b}$ and $D^{t}$
\beq
m_{b 1} D^{b}_{R 21} D^{b *}_{L 11} +m_{b 2} D^{b}_{R 22} D^{b *}_{L 12}=
-[m_{t 1} D^{t}_{R 21} D^{t *}_{L 11} +m_{t 2} D^{t}_{R 22} D^{t *}_{L 12}].
\label{condition}
\eeq
 Eq.(\ref{condition}) is an important constraint relating $D^b$ and $D^t$ which is used as a check on the
 numerical analysis. 
   
Next we  consider  the mixings of the squarks and the mirror squarks. 
We write the superpotential in terms of the scalar fields of interest as follows
\beqn
W= -\mu \epsilon_{ij} H_1^i H_2^j+\epsilon_{ij}  [h_{1}  H_1^{i} \tilde q_L ^{j}\tilde b^c_L
 +h_{1}'  H_2^{j} \tilde q_L ^{i} \tilde t^c_L
+h_{2}  H_1^{i} \tilde Q^c{^{j}}\tilde T_{L}
 +h_{2}'  H_2^{j} \tilde Q^c{^{i}} \tilde B_{L}]\nonumber\\
+ h_{3} \epsilon_{ij}  \tilde Q^c{^{i}}\tilde q_L^{j}
 + h_{4} \tilde b^c_L \tilde  B_{L}  +  h_{5} \tilde t^c_L \tilde T_{L}.
\label{superpotentials}
\eeqn
The mass$^2$ matrix of the squark - mirror squark comes from three sources, i.e., the F term,
the D term, and the soft susy breaking terms.
Using the above superpotential and after the breaking of  the electroweak symmetry we get for the
mass part of the lagrangian $ {\cal L}_F$ and $ {\cal L}_D$ the following set of terms 
\beqn
-{\cal L}_F=(m^2_B +|h_3|^2)\tilde B_R \tilde B^*_R +(m^2_T +|h_3|^2)\tilde T_R \tilde T^*_R
+(m^2_B +|h_4|^2)\tilde B_L \tilde B^*_L\nonumber\\
 +(m^2_T +|h_5|^2)\tilde T_L \tilde T^*_L
+(m^2_{b} +|h_4|^2)\tilde b_R \tilde b^*_R +(m^2_{t} +|h_5|^2)\tilde t_R \tilde t^*_R
+(m^2_{b} +|h_3|^2)\tilde b_L \tilde b^*_L\nonumber\\ +(m^2_{t} 
+|h_3|^2)\tilde t_L \tilde t^*_L
+\{-m_{b} \mu^* \tan\beta \tilde b_L \tilde b^*_R -m_{T} \mu^* \tan\beta \tilde T_L \tilde T^*_R 
-m_{t} \mu^* \cot\beta \tilde t_L \tilde t^*_R\nonumber\\
 -m_{B} \mu^* \cot\beta \tilde B_L \tilde B^*_R +(m_B h^*_3 +m_{b} h_4) \tilde B_L \tilde b^*_L 
+(m_B h_4 +m_{b} h^*_3) \tilde B_R \tilde b^*_R\nonumber\\
+(m_{t} h_5 -m_{T} h^*_3) \tilde T_L \tilde t^*_L
+(m_{T} h_5 -m_{t} h^*_3) \tilde T_R \tilde t^*_R
+h.c. \},
\eeqn
and
\beqn
-{\cal L}_D=\frac{1}{2} m^2_Z \cos^2\theta_W \cos 2\beta \{\tilde t_L \tilde t^*_L -\tilde b_L \tilde b^*_L 
+\tilde B_R \tilde B^*_R -\tilde T_R \tilde T^*_R\}\nonumber\\
+\frac{1}{2} m^2_Z \sin^2\theta_W \cos 2\beta \{-\frac{1}{3}\tilde t_L \tilde t^*_L
+\frac{4}{3}\tilde t_R \tilde t^*_R
 -\frac{1}{3}\tilde b_L \tilde b^*_L\nonumber\\
-\frac{4}{3}\tilde T_L \tilde T^*_L 
+\frac{1}{3}\tilde B_R \tilde B^*_R +\frac{1}{3}\tilde T_R \tilde T^*_R +\frac{2}{3} \tilde B_L \tilde B^*_L 
-\frac{2}{3} \tilde b_R \tilde b^*_R\}.
\eeqn
Next we add the general set of soft supersymmetry breaking terms to the scalar potential so that 
\beqn
V_{soft}=\tilde M^2_q \tilde q^{i*}_L \tilde q^i_L +\tilde M^2_{Q} \tilde Q^{ci*} \tilde Q^{ci}
+\tilde M^2_{t} \tilde t^{c*}_L \tilde t^c_L 
+\tilde M^2_{b} \tilde b^{c*}_L \tilde b^c_L +\tilde M^2_B \tilde B^*_L \tilde B_L
 + \tilde M^2_T \tilde T^*_L \tilde T_L \nonumber\\
+\epsilon_{ij} \{h_1 A_{b} H^i_1 \tilde q^j_L \tilde b^c_L 
-h'_1 A_{t} H^i_2 \tilde q ^j_L \tilde t^c_L
+h_2 A_T H^i_1 \tilde Q^{cj} \tilde T_L
-h'_2 A_B H^i_2 \tilde Q^{cj} \tilde B_L +h.c.\}.
\eeqn
From ${\cal L}_{F,D}$ and
by giving the neutral Higgs their vacuum expectation values in  $V_{soft}$ we can produce 
the mass$^2$  matrix $M^2_{\tilde b}$ for the sbottom and the  mirror sbottoms 
 in the basis $(\tilde  b_L, \tilde B_L, \tilde b_R, 
\tilde B_R)$. We  label the matrix  elements of these as $(M^2_{\tilde b})_{ij}= M^2_{ij}$ where
\beqn
M^2_{11}=\tilde M^2_q +m^2_{b} +|h_3|^2 -m^2_Z cos 2 \beta (\frac{1}{2}-\frac{1}{3}\sin^2\theta_W), \nonumber\\
M^2_{22}=\tilde M^2_B +m^2_{B} +|h_4|^2 +\frac{1}{3}m^2_Z cos 2 \beta \sin^2\theta_W, \nonumber\\
M^2_{33}=\tilde M^2_{b} +m^2_{b} +|h_4|^2 -\frac{1}{3}m^2_Z cos 2 \beta \sin^2\theta_W, \nonumber\\
M^2_{44}=\tilde M^2_{Q} +m^2_{B} +|h_3|^2 +m^2_Z cos 2 \beta (\frac{1}{2}-\frac{1}{3}\sin^2\theta_W), \nonumber\\
M^2_{12}=M^{2*}_{21}=m_B h^*_3 +m_{b} h_4,
M^2_{13}=M^{2*}_{31}=m_{b}(A^*_{b} -\mu \tan\beta),\nonumber\\
M^2_{14}=M^{2*}_{41}=0,
M^2_{23}=M^{2*}_{32}=0,\nonumber\\
M^2_{24}=M^{2*}_{42}=m_B(A^*_B -\mu \cot \beta),
M^2_{34}=M^{2*}_{43}=m_B h_4 +m_{b} h^*_3.
\eeqn

Here the terms $M^2_{11}, M^2_{13}, M^2_{31}, M^2_{33}$ arise from soft  
breaking in the  sector $\tilde b_L, \tilde b_R$. Similarly the terms 
$M^2_{22}, M^2_{24},$  $M^2_{42}, M^2_{44}$ arise from soft  
breaking in the  sector $\tilde B_L, \tilde B_R$. The terms $M^2_{12}, M^2_{21},$
$M^2_{23}, M^2_{32}$, $M^2_{14}, M^2_{41}$, $M^2_{34}, M^2_{43},$  arise  from mixing between the sbottoms  and 
the mirrors.  We assume that all the masses are of the electroweak scale 
so all the terms enter in the mass$^2$ matrix.  We diagonalize this hermitian mass$^2$ matrix  by the
 unitary transformation 
$
 \tilde D^{b \dagger} M^2_{\tilde b} \tilde D^{b} = diag (m^2_{\tilde b_1},  
m^2_{\tilde b_2}, m^2_{\tilde b_3},  m^2_{\tilde b_4})$.
There is a  similar mass$^2$  matrix in the stop sector.
In the basis $(\tilde  t_L, \tilde T_L, \tilde t_R, \tilde T_R)$ 
we can write the  mass$^2$ matrix for the stops and the mirror stops 
in the form 
$(M^2_{\tilde t})_{ij}=m^2_{ij}$ where
\beqn
m^2_{11}=\tilde M^2_q +m^2_{t} +|h_3|^2 +m^2_Z cos 2 \beta (\frac{1}{2}-\frac{2}{3}\sin^2\theta_W), \nonumber\\
m^2_{22}=\tilde M^2_T +m^2_{T} +|h_5|^2 -\frac{2}{3}m^2_Z cos 2 \beta \sin^2\theta_W, \nonumber\\
m^2_{33}=\tilde M^2_{t} +m^2_{t} +|h_5|^2 +\frac{2}{3}m^2_Z cos 2 \beta \sin^2\theta_W, \nonumber\\
m^2_{44}=\tilde M^2_{Q} +m^2_{T} +|h_3|^2 -m^2_Z cos 2 \beta (\frac{1}{2}-\frac{2}{3}\sin^2\theta_W), \nonumber\\
m^2_{12}=m^{2*}_{21}=-m_T h^*_3 +m_{t} h_5,
m^2_{13}=m^{2*}_{31}=m_{t}(A^*_{t} -\mu \cot\beta),\nonumber\\
m^2_{14}=m^{2*}_{41}=0,
m^2_{23}=m^{2*}_{32}=0,\nonumber\\
m^2_{24}=m^{2*}_{42}=m_T(A^*_T -\mu \tan \beta),
m^2_{34}=m^{2*}_{43}=m_T h_5 -m_{t} h^*_3.
\eeqn

As in the sbottom - mirror sbottom sector 
here also the terms $m^2_{11}, m^2_{13}, m^2_{31}, m^2_{33}$ arise from soft  
breaking in the  sector $\tilde t_L, \tilde t_R$. Similarly the terms 
$m^2_{22}, m^2_{24}$,  $m^2_{42}, m^2_{44}$ arise from soft  
breaking in the  sector $\tilde T_L, \tilde T_R$. The terms $m^2_{12}, m^2_{21},$
$m^2_{23}, m^2_{32}$, $m^2_{14}, m^2_{41}$, $m^2_{34}, m^2_{43},$  arise  
from mixing between the physical sector and 
the mirror sector.  Again as in the sbottom- mirror sbottom  sector 
we assume that all the masses are of the electroweak size
so all the terms enter in the mass$^2$ matrix.  This mass$^2$  matrix can be diagonalized  by the
 unitary transformation 
$
 \tilde D^{t\dagger} M^2_{\tilde t} \tilde D^{t} = diag (m^2_{\tilde t_1},  
m^2_{\tilde t_2}, m^2_{\tilde t_3},  m^2_{\tilde t_4})
$.
The physical bottom and top states are $b\equiv b_1, t\equiv t_1$,
and the states $b_2, t_2$ are heavy states with mostly mirror particle content. 
The states $\tilde b_i, \tilde t_i; ~i=1-4$ are the sbottom and stop states and their mirrors. 
For the case of  
no mixing these limits are as  follows: 
\beqn
\tilde b_1\to \tilde b_L, ~\tilde b_2\to \tilde B_L, ~\tilde b_3\to \tilde b_R, ~
\tilde b_4\to \tilde B_R,\nonumber\\
~\tilde t_1\to \tilde t_L, ~\tilde t_2\to \tilde T_L, ~\tilde t_3\to \tilde t_R, ~
\tilde t_4\to \tilde T_R.
\eeqn

The couplings $h_3$, $h_4$ and $h_5$ can be complex and thus the matrices $D^{b}_{L,R}$ and
$D^{t}_{L,R}$ will have complex elements that would produce electric dipole moments through
their arguments discussed in the text  of the paper. 
Also the trilinear couplings $A_{t, b, B, T}$ could be
complex and produce electric dipole moment through the arguments of $\tilde D^{t}$ and $\tilde D^{b}$.
We will assume for simplicity that this is the only part in the theory that has
CP violating phases.
Thus the $\mu$ parameter is considered real along
with the other trilinear couplings in the theory. 
The above allows one to automatically satisfy the constraints from the upper limits on 
 the EDMs of  the electron,  the neutron
and of  Hg and of Thallium.

\clearpage
 
\vspace{-2cm}

\clearpage


\begin{thebibliography}{999}

\bibitem{Hoogeveen:1990cb}
  F.~Hoogeveen,
  Nucl.\ Phys.\  B {\bf 341} (1990) 322;
 M.~E.~Pospelov and I.~B.~Khriplovich,
  Sov.\ J.\ Nucl.\ Phys.\  {\bf 53} (1991) 638
  [Yad.\ Fiz.\  {\bf 53} (1991) 1030].

\bibitem{Soni:1992tn}
  A.~Soni and R.~M.~Xu,
  Phys.\ Rev.\ Lett.\  {\bf 69}, 33 (1992).



\bibitem{Ibrahim:2010va}
  T.~Ibrahim and P.~Nath,
  Phys.\ Rev.\  D {\bf 81}, 033007 (2010)
  [arXiv:1001.0231 [hep-ph]].

\bibitem{Ibrahim:2008gg}
  T.~Ibrahim and P.~Nath,
  Phys.\ Rev.\  D {\bf 78}, 075013 (2008);
  [arXiv:0806.3880 [hep-ph]];
  Nucl.\ Phys.\ Proc.\ Suppl.\  {\bf 200-202}, 161 (2010)
  [arXiv:0910.1303 [hep-ph]].

\bibitem{Ibrahim:2007fb}
  T.~Ibrahim and P.~Nath,
  Rev.\ Mod.\ Phys.\  {\bf 80}, 577 (2008);
  arXiv:hep-ph/0210251;
 J.~R.~Ellis, J.~S.~Lee and A.~Pilaftsis,
  JHEP {\bf 0810}, 049 (2008)
  [arXiv:0808.1819 [hep-ph]];
 M.~Pospelov and A.~Ritz,
  Annals Phys.\  {\bf 318}, 119 (2005)
  [arXiv:hep-ph/0504231].

\bibitem{Georgi:1979md}
  H.~Georgi,
  Nucl.\ Phys.\  B {\bf 156}, 126 (1979);
  F.~Wilczek and A.~Zee,
  Phys.\ Rev.\  D {\bf 25}, 553 (1982);
J. Maalampi, J.T. Peltoniemi, and M. Roos, PLB 220, 441(1989);
  J.~Maalampi and M.~Roos,
  Phys.\ Rept.\  {\bf 186}, 53 (1990);
  K.~S.~Babu, I.~Gogoladze, P.~Nath and R.~M.~Syed,
  Phys.\ Rev.\  D {\bf 74}, 075004 (2006):
  Phys.\ Rev.\  D {\bf 74}, 075004 (2006); 
  P.~Nath and R.~M.~Syed,
  Phys.\ Rev.\  D {\bf 81}, 037701 (2010).

 \bibitem{Senjanovic:1984rw}
G.~Senjanovic, F.~Wilczek and A.~Zee,
  Phys.\ Lett.\  B {\bf 141}, 389 (1984);

\bibitem{Barger:2006fm}
  V.~Barger, J.~Jiang, P.~Langacker and T.~Li,
  Int.\ J.\ Mod.\ Phys.\  A {\bf 22}, 6203 (2007).

\bibitem{Lavoura:1992qd}
  L.~Lavoura and J.~P.~Silva,
  Phys.\ Rev.\  D {\bf 47}, 1117 (1993).

\bibitem{Maekawa:1995ha}
  N.~Maekawa,
  Phys.\ Rev.\  D {\bf 52}, 1684 (1995).

\bibitem{Morrissey:2003sc}
  D.~E.~Morrissey and C.~E.~M.~Wagner,
  Phys.\ Rev.\  D {\bf 69}, 053001 (2004)
  [arXiv:hep-ph/0308001].

\bibitem{Choudhury:2001hs}
  D.~Choudhury, T.~M.~P.~Tait and C.~E.~M.~Wagner,
  Phys.\ Rev.\  D {\bf 65}, 053002 (2002)
  [arXiv:hep-ph/0109097].

\bibitem{Liu:2009cc}
  C.~Liu,
  Phys.\ Rev.\  D {\bf 80}, 035004 (2009)
  [arXiv:0907.3011 [hep-ph]].

\bibitem{Babu:2008ge}
  K.~S.~Babu, I.~Gogoladze, M.~U.~Rehman and Q.~Shafi,
  Phys.\ Rev.\  D {\bf 78}, 055017 (2008).

\bibitem{Martin:2009bg}
  S.~P.~Martin,
  Phys.\ Rev.\  D {\bf 81}, 035004 (2010)
  [arXiv:0910.2732 [hep-ph]];
S.~P.~Martin,
  arXiv:1006.4186 [hep-ph].


\bibitem{Graham:2009gy}
  P.~W.~Graham, A.~Ismail, S.~Rajendran and P.~Saraswat,
  arXiv:0910.3020 [hep-ph].
  
  
  
\bibitem{Arnold:2010vs}
  J.~M.~Arnold, B.~Fornal and M.~Trott,
  arXiv:1005.2185 [hep-ph].


  
  

\bibitem{Jezabek:1994zv}
  M.~Jezabek and J.~H.~Kuhn,
  Phys.\ Lett.\  B {\bf 329}, 317 (1994);
  C.~A.~Nelson, B.~T.~Kress, M.~Lopes and T.~P.~McCauley,
  Phys.\ Rev.\  D {\bf 56}, 5928 (1997);
  V.~M.~Abazov {\it et al.}  [D0 Collaboration],
  Phys.\ Rev.\ Lett.\  {\bf 100}, 062004 (2008).

\bibitem{Kane:1991bg}
  G.~L.~Kane, G.~A.~Ladinsky and C.~P.~Yuan,
  Phys.\ Rev.\  D {\bf 45}, 124 (1992).

\bibitem{Schmidt:1992et}
  C.~R.~Schmidt and M.~E.~Peskin,
  Phys.\ Rev.\ Lett.\  {\bf 69}, 410 (1992).

\bibitem{Cuypers:1994ih}
  F.~Cuypers and S.~D.~Rindani,
  Phys.\ Lett.\  B {\bf 343}, 333 (1995)
  [arXiv:hep-ph/9409243].

\bibitem{Frey:1997sg}
  R.~Frey {\it et al.},
{\it In the Proceedings of 1996 DPF / DPB Summer Study on New Directions for High-Energy Physics (Snowmass 96), Snowmass, Colorado, 25 Jun - 12
Jul 1996, pp STC119}
  [arXiv:hep-ph/9704243].


\bibitem{atwood}
  D.~Atwood, S.~Bar-Shalom, G.~Eilam and A.~Soni,
  Phys.\ Rept.\  {\bf 347}, 1 (2001)
  [arXiv:hep-ph/0006032].




\bibitem{Ibrahim:2003ca}
  T.~Ibrahim and P.~Nath,
  Phys.\ Rev.\  D {\bf 67}, 095003 (2003).
  [arXiv:hep-ph/0301110].


 \bibitem{as2}
  A.~Soni and R.~M.~Xu,
  Phys.\ Rev.\ D.\  {\bf 45}, 2405 (1992).

      \bibitem{Bartl:1997wf}
        A.~Bartl, E.~Christova, T.~Gajdosik and W.~Majerotto,
        Nucl.\ Phys.\ Proc.\ Suppl.\  {\bf 66}, 75 (1998)
        [arXiv:hep-ph/9709219].

\bibitem{Hollik:1998vz}
  W.~Hollik, J.~I.~Illana, S.~Rigolin, C.~Schappacher and D.~Stockinger,
  Nucl.\ Phys.\  B {\bf 551}, 3 (1999)
  [Erratum-ibid.\  B {\bf 557}, 407 (1999)]
  [arXiv:hep-ph/9812298].

\bibitem{NovalesSanchez:2009zz}
  H.~Novales-Sanchez and J.~J.~Toscano,
  AIP Conf.\ Proc.\  {\bf 1116}, 443 (2009).

\bibitem{Huang:1994zg}
  C.~S.~Huang and T.~J.~Li,
  Z.\ Phys.\  C {\bf 68}, 319 (1995).

\bibitem{Atwood:1992vj}
  D.~Atwood, A.~Aeppli and A.~Soni,
  Phys.\ Rev.\ Lett.\  {\bf 69}, 2754 (1992).

\bibitem{Poulose:1997xk}
  P.~Poulose and S.~D.~Rindani,
  Phys.\ Rev.\  D {\bf 57}, 5444 (1998)
  [Erratum-ibid.\  D {\bf 61}, 119902 (2000)]
  [arXiv:hep-ph/9709225].

\bibitem{Choi:1995kp}
  S.~Y.~Choi and K.~Hagiwara,
  Phys.\ Lett.\  B {\bf 359}, 369 (1995)
  [arXiv:hep-ph/9506430].


\bibitem{Ibrahim:1997gj}
  T.~Ibrahim and P.~Nath,
  Phys.\ Rev.\  D {\bf 57}, 478 (1998).
  [arXiv:hep-ph/9708456].



  \bibitem{incancel}
   T.~Ibrahim and P.~Nath,
  Phys.\ Lett.\  B {\bf 418}, 98 (1998);
  Phys.\ Rev.\  D {\bf 58}, 111301 (1998);
  M. Brhlik, G.J. Good, and G.L. Kane, Phys. Rev. {\bf D59}, 115004
 (1999); A. Bartl, T. Gajdosik, W. Porod, P. Stockinger, and
 H. Stremnitzer,  Phys. Rev. {\bf 60}, 073003(1999);
 S. Pokorski, J. Rosiek and C.A. Savoy, 
 Nucl.Phys. {\bf B570}, 81(2000);
 E.~Accomando, R.~Arnowitt and B.~Dutta,
Phys.\ Rev.\ D {\bf 61}, 115003 (2000);
  U. Chattopadhyay, T. Ibrahim, D.P. Roy, Phys.Rev.D64:013004,2001;
 C.~S.~Huang and W.~Liao,
Phys.\ Rev.\ D {\bf 61}, 116002 (2000);
ibid, Phys.\ Rev.\ D {\bf 62}, 016008 (2000);
 M. Brhlik, L. Everett, G. Kane and J. Lykken, Phys. Rev.
 Lett. {\bf 83}, 2124, 1999; Phys. Rev. {\bf D62}, 035005(2000);
T.~Ibrahim and P.~Nath,
  Phys.\ Rev.\  D {\bf 61}, 095008 (2000);
Phys.\ Rev.\  D {\bf 63}, 035009 (2001);
 T. Falk, K.A. Olive, M. Prospelov, and R. Roiban, Nucl. Phys. 
 {\bf B560}, 3(1999); V.~D.~Barger, T.~Falk, T.~Han, J.~Jiang, T.~Li 
 and T.~Plehn,
Phys.\ Rev.\ D {\bf 64}, 056007 (2001);
 T.~Ibrahim and P.~Nath,
  Phys.\ Rev.\  D {\bf 61}, 093004 (2000);
T. Ibrahim,  Phys. Rev. D64 (2001) 035009.


\bibitem{lpr}
 Y.~Li, S.~Profumo and M.~Ramsey-Musolf,
  arXiv:1006.1440 [hep-ph].

\bibitem{Nath:2010zj}
See T. Ibrahim,  J. S. Lee, P. Nath, and A. Pilaftsis, 
"CP violation at the LHC", in 
  P.~Nath, B.D. Nelson {\it et al.},
  Nucl.\ Phys.\ Proc.\ Suppl.\  {\bf 200-202}, 185 (2010)
  [arXiv:1001.2693 [hep-ph]].




\end{thebibliography}
\end{document}